\documentstyle[12pt,aasms4]{article}

\begin{document}
\pagenumbering{arabic}

\title{DIAMETERS OF OPEN STAR CLUSTERS}

\author{Sidney van den Bergh}
\affil{Dominion Astrophysical Observatory, Herzberg Institute of Astrophysics, National Research Council of Canada, 5071 West Saanich Road, Victoria, BC, V9E 2E7, Canada. email: sidney.vandenbergh@nrc.gc.ca}

\begin{abstract}

The present paper presents a tabulation of data on all 600 Galactic open clustres for which it is presently possible to calculate 
linear diameters. As expected, the youngest ``clusters'' 
with ages  $<$ 15 Myr,  contain a significant ($\geq$ 20\%) admixture 
of associations. Among intermediate-age clusters, 
with ages in the range 15 Myr to 1.5 Gyr, the median cluster 
diameter is found to increase with age. Small compact clusters are 
 rare among objects with with ages $>$ 1.5 Gyr. Open clusters with ages $>$ 1 Gyr appear to form  what might be termed a ``cluster thick disk'', part of which consistst of objects that were probably captured 
gravitationally by the main body of the Galaxy.  

\end{abstract}


\section{INTRODUCTION}

      Star clusters may be classified on the basis of three 
objective parameters: mass, age and size (Whitmore 2004). These 
parameters are now available for the majority of Galactic globular clusters 
(e.g. van den Bergh \& Mackey 2004, Mackey \& van den Bergh 2005), 
and for some of the globulars in nearby galaxies that are members 
of the Local Group. However, such basic information is still 
lacking for the majority of Galactic open star clusters.

   For very  early discussions of Galactic open clusters, and  compilations 
of data  on these objects,the reader is refereed  to ten Bruggencate (1927) and Sawyer-Hogg (1959). More recently Lyng\aa (1987) has published an 
updated catalog of data on open clusters. Using this 
information Janes, Tilley \& Lyng\aa (1988) discused some of the general properties of the Galactic open cluster system. A much more 
detailed compilation of data on Galactic open cluster has recently 
been published  by Dias et al. (2002). A version of this catalog 
(Dias \& L\'{e}pine 2005), that had been updated to March 2005, was 
used for our investigation. A statistical discussion of some
of this material has very recently been published by Bonatto et al. (2005). It is the purpouse of the present investigation to assemble all presently available data on those  Galactic clusters  for which linear diameters could be calculated from the published data. It is hoped that the (often only
fragmentary) information that is presently available might  provide some  
useful constraints on the evolutionary history of the system or Galactic open 
clusters, or  perhaps even on that of the  Milky Way galaxy itself. 
In particular it is hoped that the large number of open clusters 
for which diameters are now available can, at lease in a statistical sense,  
compensate for the relatively low quality of some of the published data on individual Galactic star clusters.

\section{THE CATALOG}

The data by Dias \& L\'{e}pine were used to compile Table 1. Only 
those clusters for which the data allowed the calculation of the 
cluster diameter are included in this table. For each cluster the 
following information is listed: (1) a cluster running number, (2) 
the cluster name, (3) the truncated Galactic longitude, (4) the 
cluster diameter in pc, (5) the distance from the Sun to the cluster 
in pc, (6) the distance of the cluster from the Galactic plane 
calculated by assuming that {\it z = R sin b}, (7) the reddening {\it E(B-V)} of the cluster as determined from photoelectric photometry, and 
(8) the logarithm of the cluster age in years, as derived from its 
dereddened color-magnitude diagram. Per chance the total number of 
clusters in the database, for which linear diameters could be 
determined, was exactly 600. A recent study by Joshi (2005) 
suggests that the present cluster data are probably more-or-less 
complete out to a distance of two or three kpc. 

\section{DISCUSSION}

\subsection{Dependence on Galactic longitude.}

The longitude distribution of the open clusters listed
in Table 1 is plotted in Figure 1. This distribution is seen to exhibit maxima in the regions of active star formation at {\it l} $\sim 125^{\circ}$ (Cassiopeia), {\it l} $\sim$ $205^{\circ}$ (Monoceros), {\it l} $\sim$ $240^{\circ}$ (Canis Major) and {\it l} $\sim$ $285^{\circ}$ (Carina). The deepest minimum occurs in the obscured region at {\it l} $\sim$ $50^{\circ}$ (Sagitta). The distribution of clusters in Galactic longitude depends on age (Dias \& L\'{e}pine 2005). The present data show that the youngest star clusters with ages $<$ 1 $ \times 10^{7}$ years  are strongly concentrated in the Carina arm $(280^{\circ} < l < 300^{\circ}$) and along the Sagittarius arm ({\it l} = $330^{\circ}$ to {\it l}  = $20^{\circ}$). On the other hand the oldest  open clusters with ages $> 1 \times 10^{9}$ years clearly favor the region with $190^{\circ} < {\it l} < 260^{\circ}$. Possible contributors to this excess of very old clusters in the zone from {\it l} $\sim$ $190^{\circ}$ to {\it l} $\sim$  $260^{\circ}$ are: (1) relatively low Galactic absorption, (2) the fact that destructive tidal forces exerted by giant molecular clouds are smallest in the Galactic anti-center direction (van den Bergh \& McClure 1980), and (3) the possible presence of a number of clusters that may be associated with the Canis Major dwarf galaxy (e.g. Bellazzini et al. 2005), rather than with the Milky Way System itself. According to Bellazzini et al. the following clusters in Table 1 are possibly associated with the CMa system: Tombaugh 2, Arp-Madore 2, NGC 2243, Melotte 66, van den Bergh-Hagen 66 and Sauer 2. All of these objects, except vdB-H66 (for which log T = 8.9) have ages greater than 1.0 Gyr.  It is not yet clear to which extent the overdensity of old clusters having (190$^{\circ} < l < 260^{\circ}$) is due to the putative Moneceros Ring in the background of the Canis Major dwarf galaxy (Conn et al. 2005), or perhaps to the ``Galactic Anticenter Structure'' discussed by Frinchaboy et al. (2004, 2005) and Martin et al. (2005).

\subsection{Distribution of cluster diameters.}

 The distribution of the diameters of the open clusters that are listed in Table 1 shows a peak at {\it D} $\sim$ 2.6 pc, with half of all clusters having {\it D} $<$ 3.5 pc. The distribution of the the diameters of clusters having different ages is plotted in Figure 2 and is listed in Table 2. The distribution of diameters of young clusters with ages $<$ 15 Myr (Fig. 2a) shows a much broader wing towards large radii than does the distribution of intermediate-age clusters having ages in the range 15 - 150 Myr (Figure 2b) . A Kolmogorov- Smirnov test shows that there is only a 0.3\% probability that these two distributions were drawn from the same parent population.  As previously noted by Janes et al. (1988) [See their Figure 3] the 
reason for the broad wing in the diameter distribution of young 
``clusters'' is probably that many of the listed objects with large 
radii are actually positive energy expanding associations 
(Ambartsumian 1954, Blaauw 1964), rather than negative energy 
bound clusters. Of the young clusters  with ages T $<$ 15 Myr listed 
in Table 2, 29\% have {\it D} $>$ 7.0 pc, compared to only 9\% of the clusters with ages in the range 15 Myr $<$ T $<$ 150 Myr. This 
suggets that $\sim$20\% of the  young ``clusters'' listed in Table 1 are actually expanding associations with {\it D} $>$ 7.0 pc rather than stable 
clusters. Since very young associations might not yet have had time to expand to {\it D} = 7 pc the total fraction of positive energy systems in the present data sample must actually be even greater than 20\%.
  
A comparison between the diameter distributions of 
intermediate-age clusters with ages in the range 15 Myr to 150 Myr 
(Figure 2b) and older clusters with ages of 150 Myr to 1.5 Gyr 
(Figure 2c) shows that the older clusters are systematically larger 
than the intermediate-age clusters. A K-S test shows that there is 
only a 0.6\% probability that these two distributions were drawn 
from the same parent population. Intermediate-age clusters are seen 
to have a peak in their diameter distribution at {\it D} $\sim$2.0 pc, compared 
to a peak diameter of {\it D} $\sim$3.0 pc for the older clusters. Possibly this increase in cluster diameter with age (which was also found by Bonatto et al. 2005) is, at least in part, due to 
the loss of gas by evolving stars (Schweitzer 2004).  Finally 
Figure 2d shows that the most ancient open clusters with 
ages $>$ 1.5 Gyr, are systematically larger than the clusters that 
have ages in the range 150 Myr -1.5 Gyr. A K-S test shows that 
there is only a 0.3\% probability that the open clusters with ages 
150 Myr to 1.5 Gyr and those with ages $>$ 1.5 Gyr were drawn from 
the same parent population. This result is somewhat 
counterintuitive because one might perhaps  have expected the 
most tightly bound clusters to survive longest. This effect is, at leaset in part, due to the fact that the oldest clusters tend to be located in the Galactic anti-center direction.

\subsection{Cluster diameter and Galactocentric distance.}

   Many years ago van den Bergh \& Morbey (1984) showed that the half-light radii of globular clusters grow significantly with increasing Galactocentric distance. This raises the question wheather a similar relation exists between  the radii and the Galactocentric distances of open clusters. Unfortunately it turns out that this question cannot yet  be answered unambiguously with the present data. The reason for this is that the apparent cluster radii listed in Table 1 are affected by the stellar background density on which a cluster is projected. In the direction of the Galactic center ({\it l} = $320^{\circ}$ to ~{\it l} = $40^{\circ}$)  low-lattitude clusters with $\mid$ b $\mid$ $ < 2.0^{\circ}$ (which are projected on high-density star
fields) are seen to appear significantly smaller than do the clusters with $\mid$ b $\mid$ $ > 2.0^{\circ}$ that appear projected on lower density star fields. A similar effect is observed in the anti-center direction ($140^{\circ} < 1 < 220^{\circ}$) where open clusters with $\mid$ b $\mid$ $< 0.6^{\circ}$ are seen to be significantly smaller than those that occur at higher latitudes. As 
a result of this selection effect that depends on stellar background (or foreground) density, it is not yet clear if the observed increase in mean cluster diameter with Galactocentric radius is intrinsic, 
or whether it might (at least in part) be due to observational 
selection effects. Such selection effects could be greatly reduced (or eliminated) by measuring the half-light radii $R_{h}$ of a significant number of Galactic open clusters. With such data on cluster $R_{h}$ values it should be possible to establish beyond reasonable doubt if  remote open clusters are  systematically larger than those that occur closer to the Galactic center. Such an effect might be expected because  disruptive tidal forces are smallest at large Glactocentric radii. Furthermore,  clusters at large Galactocentric distances are less likely to be destroyed by 
interactions with giant giant molecular clouds (van den Bergh \& 
McClure 1980). 

 There is a possible inherent problem with the present data which results from the fact that unusually rich clusters, such as 
NGC 6791 ({\it D} = 17 pc), and NGC 7789 ({\it D} = 17 pc,  may have been 
traced to large radii as a result of their huge stellar population, 
rather than because of an intrinsically large half-light radius. 
Clearly it would be important to measure the half-light radii of 
Galactic open clusters so as to avoid this type of bias for the 
cluster radii that have been published in the literature. Finally 
the diameters derived from a Hipparcos proper motion search for 
clusters and associations by Platais et al. (1998) are almost all 
unusually large. 

\subsection{Distribution of cluster distances from the Galactic plane.}

  Figure 3 shows that the open clusters in Table 1 are 
strongly concentrated towards the Galactic plane, with half of 
all clusters having $\mid$ z $\mid$ $<$ 48pc. Some caution should, however be 
used in analyzing and interpreting such data because (Joshi 2005, Reed 2005). the frequency of nearby open clusters sppears to peak  
at {\it z} $\sim$ -25 pc. It is not immediately clear how much of this offset 
is due to a displacement of the local dust layer from the
Galactic plane, the location of the Sun above the Galactic 
plane and (or) an asymmetry in the distribution of young star 
formation near the Sun with respect to the Galactic plane.  The 
distribution of Galactic open clusters in $\mid$ z $\mid$ exhibits a broad 
tail containing 13 objects (see Table 4) that are located at more 
more than 1.0 kpc from the Galactic plane. It is of interest to 
note that all such objects, for which it has been possible to 
derive ages from color-magnitude diagrams, are older than $1 \times 10^{9}$ years. Perhaps the most famous example of 
such very old clusters at high $\mid$ z $\mid$ is the metal-rich open cluster 
NGC 6791 (Stetson, Bruntt \& Grundahl 2003, King et al. 2005) 
which is situated at {\it z} = +1.1 kpc. It is of particular interest 
to note that four of the 13 objects in Table 4, that are situated 
far from the Galactic plane, appear to be associated with the 
Canis Major dwarf system. This suggests that captured dwarf 
galaxies may have provide a  significant contribution 
to the population of open clusters with $\mid$ z $\mid$ $>$ 1 kpc. One 
might perhaps think of such open clusters far from the 
Galactic plane as constituting a kind of ``thick disk''  [cf. Dalcanton, 
Seth \& Yoachim 2005), i.e. a population that has (at least in
part) been derived from tidal capture of initially extragalactic objects The data in Table 1 clearly show that the oldest clusters are situated 
at greater distances from the Galactic plane than are the more 
recently formed clusters. Of the 44 clusters with ages $>$ 1.5 Gyr 
half are located at $\mid$ z $\mid$ $<$ 278 pc, compared to half having $\mid$ z $\mid$ $<$ 48 pc for the entire cluster sample. This large difference is, no 
doubt, due to both (1) preferential destruction of old clusters 
close to the Galactic plane by giant molecular clouds and disk 
shocks, and (2)  to the fact that the oldest open cluster sample is 
more enriched in objects that were gravitationally  captured by 
the Galaxy. 

It is also noted that the vast majority of clusters located 
towards the Galactic center (${\it l} = 320^{\circ}$  to {\it l} =  $40^{\circ})$ have $\mid$ z $\mid$  $<$ 100 pc, wheas most of the clusters in the anti-center direction ($140^{\circ} < {\it l} < 220^{\circ}$) are situated at $\mid$ z $\mid$ $>$ 100 pc.

\subsection{Distribution of reddening values.}

 The distribution of reddening values for the clusters in
Table 1 is listed in Table 5. The observed cluster reddenings range from quite low values for nearby clusters, and for objects at high Galactic latitudes, to {\it E(B-V)} = 2.25 and {\it E(B-V)} = 3.00 for the highly obscured low latitude clusters van den Bergh - Hagen No. 245 and Westerlund No. 1, respectively. Only $\sim$ 2\% of the clusters in the present sample have {\it E(B-V)} $>$ 1.50. Most of these highly reddened clusters are situated within 30$^{\circ}$ of the Galactic center. On the other hand the majority of the little reddened clusters are located in the direction towards the Galactic anti-center. A detailed discussion of the distribution of reddening values from a slightly larger sample of 722 clusters (which did not all meet the requirement that they have published diameters) has recently been given by Joshi (2005). His reddening map appears to show evidence for a $\sim$ 4 kpc long dust arm that comes as close as $\sim$ 1.5 kpc at l $\sim 4 0^{\circ}$.

\subsection{Distribution of cluster ages.}

 Ages are available for 586 of the open clusters listed in Table 1. These clusters are found to have ages that range from $1 \times 10^{6}$ years (NGC 6618) to $1.2 \times 10^{10}$ years (Berkeley 17). The distribution of observed open cluster ages is given in Table 6 and is shown in Figure 5. These data, and those for the slightly larger sample of Joshi (2005), both exhibits a broad age maximum centered at log {\it T} $\sim$ 8.0 on which a much narrower subsidiary peak at log {\it T} = 7.1 (12 million
years) appears to be superposed. This narrow age peak seems
to be mainly associated with groupings of young clusters at longitudes $120^{\circ}-140^{\circ}, 230^{\circ}-250^{\circ}$ and $280^{\circ}-300^{\circ}$. The observed peak at log  {\it T} $\sim$ 8.0 is in good agreement with theoretical expectations  (Wielen 1971)

\section{Desiderata}

   For a variety of reasons far less structural information 
is available on Galactic open clusters than is the case 
for Galactic globulars. It would be particularly valuable 
to obtain a homogeneous set of half-light diameters for 
a large sample of open clusters, to see how this parameters 
depends on age, Galactocentric distance etc. Furthermore
it might be interesting to look for possible Galactic progenitors to objects like the ``faint fuzzies'' of Larsen \& Brodie (2000). Due to heavy  contamination by disk stars it 
would be particularly important to use two-color photometry
to filter out field stars before attempting to measure such 
cluster half-light radii. A telescope of only moderate aperture would be required to make these important measurements. It would also be important to obtain homogeneous photometric data on the integrated photometric properties of a representative sample of Galactic open clusters.

\section{CONCLUSIONS}

  It is confirmed  that young ``clusters'' with ages $<$ 15 Myr contain 
a  significant  sub population of objects with diameters $>$ 7 pc. 
The majority of these large structures are, as was previously noted 
by Janes et al. (1988), probably positive energy expanding 
associations, rather than stable negative energy star clusters. 
Over the age range from 1.5 Myr to 1.5 Gyr the diameters of 
clusters are found to increase with age. Small compact clusters 
are conspicuously absent among the oldest objects with 
ages $>$ 1.5 Gyr. The radii of open clusters in the direction 
of the Galactic center appear to be significantly smaller 
than those of clusters seen in the anti-center direction. In 
this respect open clusters resemble globular clusters, which 
are well known to be biggest at large Galactoentric distances. 
This apparent dependence of cluster radius on Galactocentric distance 
might be caused by (1) the preferential destruction of large 
clusters by disk/bulge shocks at small values of  R$_{gc}$,  (2) 
the paucity of destructive interactions with giant molecular 
clouds at large values of R$_{gc}$, and (3) observational selection effects resulting from the high stellar background density in the direction  towards the Galactic center.

I thank Luc Simard and Michael Peddle for their help 
with the figures,  Brenda Parrish for her assistance with the 
manuscript, and Ken Janes for his very helpful referee report. 

\newpage

\centerline{\bf REFERENCES}

\noindent  Ambartsumian, V. A. 1954, I. A. U. Transactions, 8, 665
 
\noindent  Bellazzini, M., Ibata, R., Martin, N., Lewis, G. F., Conn, B., \& Irwin, M. J. 2005, MNRAS (in press = astro-ph/0504494)

\noindent  Blaauw, A. 1964, ARAA, 2, 213

\noindent Bonatto, C., Kerber, L. O., Bica, E. \& Santiago, B. X. 2005, A\&A (in press = astro-ph/0509804)

\noindent Conn, B. C., Martin, N. F., Lewis, G. F., Ibata, R. A., Bellazzini, M, \& Irwin, M. J. 2005, MNRAS, 362, 475

\noindent Dalcanton, J. J., Seth, A. \& Yoachim, P. 2005 astro-ph/0509700

\noindent  Dias, W. S., Alessi, B. S., Moitinho, \& L\'{e}pin, J. R. D. 2002, A\&A, 389, 871

\noindent Dias, W. S., \& L\'{e}pine, J. R. D. 2005 A\&A in press = astro-ph/0503083

\noindent Frinchaboy, P. M., Majewski, S. R., Crane, J. D., Reid, I. N., Rocha-Pinto, H. J., Phelps, R. L., Patterson, R. J. \& Munoz, R. R. 2004, ApJ, 602, L21 

 \noindent Frinchaboy, P. M., Munoz, R. R., Phelps, R. L., Majewski, S. M. \& Kunkel, W. E. 2005 astro-ph/0509742 

\noindent Janes, K. A., Tilley, C.  \& Lyng\aa 1988, AJ, 95, 771

\noindent Joshi, Y. C. 2005, MNRAS in press = astro-ph/0507069

\noindent King, I. R., Bedin, L. R., Piotto, G., Cassisi, S. \& Anderson, J. 2005 AJ (in press = astro-ph/0504627)

\noindent Larsen, S. S. \& Brodie, J. P. 2000, AJ, 120. 2938

\noindent Lyng\aa , G. 1887, Computer Based Catalogue of Open Cluster Data 5th Edition (Strasbourg:CDS)

\noindent Mackey, A. D. \& van den Bergh, S. 2005 MNRAS, 360, 631

\noindent Martin, N. F., Ibata, R. A., Conn, B. C., Lewis, G. F., Bellazzini, M. \& Irwin, M. J. 2005, MNRAS, 362, 906

\noindent Platais, I., Kozhurina-Platais, V. \& van Leeuwen, F. 1998 AJ, 116, 2423

\noindent  Reed, B. C. 2006, JRASC (submitted = astro-ph/0507655)

\noindent Sawyer-Hogg, H. 1959, Handbuch der Physik, 53, 129

\noindent Schweizer, F. 2004, in The Formation and Evolution of Massive Young Star Clusters = ASP Conference Series No. 322, Eds. H.J.G.L.M. Lamers, L. J. Smith and A. Nota, ASP: San Francisco),p.111

\noindent Stetson, P. B., Bruntt, H. \& Grundahl, F. 2003, PASP, 115, 413 

\noindent ten Bruggencate, P. 1927 Sternhaufen (Berlin: Springer)

\noindent van den Bergh, S., \& Mackey, A. D. 2004 MNRAS, 354, 713
 
\noindent van den Bergh, S., \& McClure, R. D. 1980 A\&A, 80, 360 

\noindent van den Bergh, S. \& Morbey, C. L. 1984, Astron. Express, 1, 1 

\noindent  Whitmore, B. C. 2004 in The Formation and Evolution of Massive Young Star Clusters = ASP Conference Series No. 322, Eds. H.J.G.L.M.Lamers, L. J. Smith and A. Nota (ASP: San Francisco), p.419

\noindent Wielen, R. 1971, A\&A 13, 309

\newpage
\centerline{\bf CAPTIONS TO FIGURES}

1. The distribution of Galactic clusters in longitude exhibits maxima in regions of active star formation at l $\sim  125^{\circ}$ (Cassiopeia), l $\sim 205^{\circ}$ (Monoceros), l $\sim 240^{\circ}$ Canis  Major) and l $\sim 285^{\circ}$ (Carina). The deepest minimum occurs near the obscured region at l $\sim 50^{\circ}$ (Sagitta).

2. Distribution of the diameters of Galactic open clusters. The youngest clusters (Fig. 2a) are seen to exhibit a large-diameter wing that is probably due to the inclusion of associations in the cluster data. Older clusters are also seen to be larger than younger ones.

3. Integral distribution of log $\mid$ z $\mid$ values for Galactic open clusters.

4. Relation between age and diameter of clusters. The Figure appears to show the following: (1) The median cluster diameter grows with increasing age. (2) Compact clusters are rare among very old clusters with ages $>$ 1.5 Gyr. Finally (3), the youngest objects with ages $<$ 15 Myr are probably a mixture of compact bound clusters and diffuse expanding associations.

\nopagebreak

\begin{deluxetable}{llllllll}
\tablewidth{0pt}  
\tablecaption{Diameters of Open Clusters} \tablehead{\colhead{No.} & \colhead{Name}  & \colhead{\it l}  & \colhead{{\it D}(pc)}  & \colhead{{\it R}(pc)}  & \colhead{{\it Z}(pc)} & \colhead{{\it E (B-V)}} & \colhead{log {\it T}}}

\startdata
                     
001 & Trumpler 31 &  002$^{\circ}$ &  1.43 &   986 &  -0039 &   0.35 & 8.87 \\
002 &  N6520      &  002 &  2.29 &  1577 &  -0078 &   0.43 & 7.72 \\
003 & N6530       &  006 &  5.42 &  1330 &  -0031 &   0.33 & 6.87 \\
004 & Bochum 14   &  006 &  0.34 &  578  &  -0005 &   1.51 & 7.00 \\
005 & N6514       &  007 &  6.65 &  816  &  -0004 &   0.19 &  7.37 \\
\nodata &  \nodata  & \nodata & \nodata  & \nodata & \nodata & \nodata & \nodata

\enddata
\it Complete table available on-line as a machine-readable table. \end{deluxetable}

\begin{deluxetable}{lcccc}
\tablewidth{0pt} 
\tablecaption{Normalized frequency distribution of Galactic cluster diameters as a function of cluster age {\it T}.} 

\tablehead {\colhead{.} & \colhead{{\it T}$<$15Myr} & \colhead{15$<${\it T}$<$150Myr}  & \colhead{0.15$<${\it T}$<$1.5Gyr}  & \colhead{{\it T}$>$1.5Gyr}\\ \colhead{{\it D}(pc)} & \colhead{N({\it D})/127}  & \colhead{N({\it D})/219}  & \colhead{N({\it D})/191}  & \colhead{N({\it D})/42}}

\startdata

 0.5  &   0.055   &    0.037  &    0.021   &   0.000\\
 1.5  &  0.157    &   0.233   &   0.136    &  0.071\\
 2.5  &  0.189    &   0.237   &   0.194   &   0.095\\
 3.5  &  0.134    &   0.169   &  0.194    &  0.143\\
 4.5  &  0.087    &   0.114   &   0.147    &  0.071\\
 5.5  &  0.055    &   0.073   &   0.079    &  0.190\\
 6.5  &  0.031    &    0.046  &    0.052  &    0.048\\
 7.5  &  0.047   &    0.014   &   0.058   &   0.095\\
 8.5  &  0.047     &  0.009   &   0.021   &   0.048\\
 9.5  &  0.016    &   0.009  &    0.031    &  0.024\\
10.5  &  0.039    &   0.005   &   0.016   &   0.024\\
11.5  &  0.016    &   0.005   &   0.010    &  0.024\\
12.5  &  0.031    &   0.000   &   0.010   &   0.048\\
13.5  &  0.016    &   0.000   &   0.000    &  0.024\\
14.5  &  0.016    &   0.009   &   0.005   &   0.000\\
$>$15 &  0.063    &   0.041   &   0.026    &  0.095\\

\enddata
\end{deluxetable}

\begin{deluxetable}{lclr}
\tablewidth{0pt}        
\tablecaption{Distribution of the values of log $\mid$ {\it z} $\mid$ in Table 1} 

\tablehead{\colhead{log $\mid$ {\it z} $\mid$} & \colhead {n(log $\mid$ {\it z} $\mid$)}  & \colhead{log $\mid$ {\it z} $\mid$}  & \colhead{n(log $\mid$ {\it z} $\mid$)}}

\startdata

\nodata    ~~~ $<$0.0   &    3         &  1.8 $\leq ~~~ <$ 2.0        &  85\\
0.0 $\leq  ~~~ <$0.2    &   5          & 2.0  $\leq  ~~~ <$2.2        &  65\\
0.2 $\leq  ~~~ <$0.4    &   11         & 2.2  $\leq  ~~~ <$2.4        &  36\\
0.4 $\leq  ~~~ <$0.6    &   10         & 2.4  $\leq  ~~~ <$2.6        &  26\\
0.6 $\leq  ~~~ <$0.8    &   13         & 2.6  $\leq  ~~~ <$2.8        &  16\\
0.8 $\leq  ~~~ <$1.0    &   29         & 2.8  $\leq  ~~~ <$3.0        &  9\\
1.0 $\leq  ~~~ <$1.2    &   43         & 3.0  $\leq  ~~~ <$3.2        &  10\\
1.2 $\leq  ~~~ <$1.4    &   71         & 3.2  $\leq  ~~~ <$3.4        &   3\\
1.4 $\leq  ~~~ <$1.6    &   81         & ~~~~ $\geq  ~~~ <$3.4        &   0\\
1.6 $\leq  ~~~ <$1.8    &   84
\enddata
\end{deluxetable}

%\newpage

\begin{deluxetable}{lrl}
\tablewidth{0pt}        
\tablecaption{Open Clusters that are located at $>$ 1.0 kpc from the Galactic plane}

\tablehead{\colhead{Name} & \colhead {{\it z}}  & \colhead{log {\it T}}}

\startdata

NGC 6791      &       +1107   &  9.64\\
NGC 7772      &       -1047   &  9.17\\
Berkeley 29   &       +2076   &  9.02\\
NGC 2420      &       +1036   &  9.05\\
Berkeley 22   &       -1073   &  9.03\\
Berkeley 20   &       -2494   &  9.78\\
Sauer 1       &       +1425   &  9.85\\
Tombaugh 2    &       -1588   &  9.01 \tablenotemark{a}\\
NGC 2243      &       -1379   & 9.03 \tablenotemark{a}\\
Arp-Madore 2   &      -1366   &  9.34 \tablenotemark{a}\\
Melotte 66     &      -1061   &   9.44 \tablenotemark{a}\\
ESO 092-18     &      -1229   &  9.02\\
v.d.Bergh-Hagen 176 & +1008   &   9.08  \tablenotemark{b}\\

\tablenotetext{a}{Probably associated with the Canis Major dwarf}
\tablenotetext{b}{Age from Frinchaboy et al. (2005)}

\enddata
\end{deluxetable}

% \newpage

\begin{deluxetable}{ll}
\tablewidth{0pt}        
\tablecaption{Distribution of cluster reddening values \it E(B-V)} 
\tablehead{\colhead{\it E(B-V)} & \colhead {\it (N)}}

\startdata

0.00 - 0.19   & 145\\
0.20 - 0.39  &  143\\
0.40 - 0.59  &  141\\
0.60 - 0.79  &   75\\
0.80 - 0.99   &  36\\
1.00 - 1.19   &  16\\
1.20 - 1.39   &  17\\
1.40 - 1.59   &   5\\
$\geq 160$   &   9\\
   ?         &   13\\

\enddata
\end{deluxetable}

\begin{deluxetable}{llrr}
\tablewidth{0pt}        
\tablecaption{Distribution of clusters ages.} 
\tablehead{\colhead{Log {\it T} (years)} & \colhead {\it N} & \colhead {Log {\it T} (years)} & \colhead {\it N }} 

\startdata

6.00 - 6.19  &   1   &  8.00 - 8.19  &   49\\
6.20 - 6.39  &  1    &  8.20 - 8.39   &   47\\
6.40 - 6.59  &  1    &  8.40 - 8.59  &   50\\
6.60 - 6.79  &  16   &   8.60 - 8.79  &   27\\
6.80 - 6.99  &  42   &   8.80 - 8.79  &   31\\
7.00 - 7.19  &  70   &   9.00 - 9.19  &   34\\
7.20 - 7.39  &  32   &   9.20 - 9.39  &   15\\
7.40 - 7.59  &  37   &   9.40 - 9.59  &   11\\
7.60 - 7.79  &  45   &   9.60 - 9.79  &   13\\
7.80 - 7.99  &  59   &   9.80 - 9.99  &    4\\
             &        &~~10.00 - 10.19   &  1\\

\enddata
\end{deluxetable}

\end{document}